\documentclass[letterpaper]{siamltex}
\usepackage{amsmath}
\usepackage{amssymb}

\DeclareMathOperator{\cov}{cov}
\DeclareMathOperator{\cum}{cum}

\newcommand{\old}[1]{{}}

\title{Large Eddy Simulation, Turbulent Transport
and the Renormalization Group\footnotemark[1]}

\author{J. Glimm\footnotemark[2] \ \footnotemark[4]
\and B. Plohr\footnotemark[3] \ \footnotemark[5]
\and D. H. Sharp\footnotemark[3] \ \footnotemark[6]}

\begin{document}
\maketitle

\renewcommand{\thefootnote}{\fnsymbol{footnote}}
\footnotetext[1]{\today}
\footnotetext[2]{Department of Applied
Mathematics and Statistics, Stony Brook University, Stony Brook,
NY 11794-3600, USA
and Computational Science Center,
Brookhaven National Laboratory, Upton, NY
11793-6000, USA}
\footnotetext[3]{Theoretical Division,
Los Alamos National Laboratory,
Los Alamos, NM, 87545, USA}
\footnotetext[4]{Supported in part by
Leland Stanford Junior University 2175022040367A
(subaward with DOE as prime sponsor), Army Research Office W911NF0910306.
This manuscript has been co-authored by
Brookhaven Science Associates, LLC, under Contract No. DE-AC02-98CH1-886 with
the U.S.~Department of Energy. The United States Government retains, and the
publisher, by accepting this article for publication, acknowledges, a
world-wide license to publish or reproduce the published form of this
manuscript, or allow others to do so, for the United States Government
purposes. Los Alamos National Laboratory Preprint Number LA-UR-12-26149}
\footnotetext[6]{Supported by Los Alamos National Laboratory (retired).}
\renewcommand{\thefootnote}{\arabic{footnote}}

\begin{abstract}
In large eddy simulations, the Reynolds averages of nonlinear terms are not
directly computable in terms of the resolved variables and require a closure
hypothesis or model, known as a subgrid scale term.
Inspired by the renormalization group (RNG),
we introduce an expansion for the unclosed terms,
carried out explicitly to all orders.
In leading order,
this expansion defines subgrid scale unclosed terms,
which we relate to the dynamic subgrid scale closure models.
The expansion, which generalizes the Leonard stress for
closure analysis, suggests a systematic higher order determination of the
model coefficients.

The RNG point of view sheds light on the nonuniqueness of the infinite
Reynolds number limit. For the mixing of $N$ species,
we see an $N+1$ parameter family of infinite
Reynolds number solutions labeled by dimensionless parameters of the
limiting Euler equations, in a manner intrinsic to the RNG itself.
Large eddy simulations, with their Leonard stress and dynamic subgrid models,
break this nonuniqueness and predict unique model coefficients on the
basis of theory. In this sense large eddy simulations 
go beyond the RNG methodology, which does not in general
predict model coefficients.
\end{abstract}



\pagestyle{myheadings}
\thispagestyle{plain}
\markboth{J. Glimm, B. Plohr, and D. H. Sharp}
{LES, Turbulent Transport, and the RNG}

\section{Introduction}
\label{sec:introduction}

\subsection{Turbulence and large eddy simulations}
\label{sec:background}
Turbulence is one of the major unsolved problems of classical
physics, a view attributed to Heisenberg, among others.
Turbulence is an instability of fluid flow which occurs at
high Reynolds numbers $\mathrm{Re} = VL/\nu$, where $V$ and $L$ are
representative velocities and lengths and $\nu$ is the kinematic
anisotropic viscosity. Many turbulent flows, including ones arising in
oceanography, atmospheric sciences, aeronautics, astrophysics
and chemical processing occur at elevated, nearly infinite Reynolds numbers.

Turbulence is a strong coupling theory with no natural length scales.
For this reason numerical solutions of turbulent flow, necessarily
cut off at some grid level, encounter difficulties in the coupling
between the resolved (grid level) data and the unresolved (subgrid level) data.
Because of the strong coupling, scale invariant nature
of subgrid scale turbulence, methods to treat
this effect on the resolved scales have been introduced.
Large Eddy Simulation (LES) is such a numerical method
in which some, but not all, of the turbulent scales are resolved, and
subgrid scale models emulate the influence of the unresolved on the resolved
scales.

The strong coupling
problem originates in the nonlinearity
of the compressible Navier-Stokes equations,
which govern the turbulent flow field. The averaging operation used
to define grid level quantities suitable for numerical computation,
when applied to nonlinear terms in the Navier-Stokes equations,
leads to new unknown flow quantities,
called subgrid scale (SGS) terms.
Supplying approximations to the SGS terms
is a process known as \emph{closure}.
The averaged nonlinear terms, before approximation, are called unclosed.

\subsection{RNG applied to LES}
\label{sec:rng-les}
In the limit $\mathrm{Re} \to \infty$, the
theory simplifies and scaling laws apply, the most famous of which
is that of Kolmogorov~\cite{Kol41}.
The renormalization group (RNG) is a method for the systematic
study of scaling laws.
The RNG framework
is explained in~\cite{McC04},
with the goal of finding 
an expression for the
unclosed terms in a LES mentioned briefly.
Our expansion accomplishes this goal, using the RNG expansion but
not an RNG fixed point. 
Relevant to the present work, we mention
our own studies
of turbulent mixing~\cite{LimIweGli10,LimKamYu12,KamLimYu11,MelRaoKau12a}.
We also mention our related K41 based existence
theorem for $L^p$ solutions of the incompressible constant density
Euler equations~\cite{CheGli12}.

The study of SGS terms is central to this paper.
Each closed SGS approximation
term is regarded as a product of a coefficient and a solution functional of
specified form. We call the latter a model. 
The RNG expansion derived here, through the dynamic method
of coefficient selection, uniquely specifies the coefficient once the model
is given. This determination is based on properties of the resolved
scales, connected to analogous properties of the unresolved scales by an
asymptotic assumption.
Thus the coefficient is determined by
theory, leaving the model selection open as a subject for research.
The search for alternate frameworks for the choice of SGS terms 
is an active topic of research, which we do not attempt to review here,
but we do discuss  problems which arise if the SGS models are
omitted completely,
as commonly occurs when the Implicit Large Eddy Simulation (ILES) method
is used in practice.

Our first main result is a closed form
expansion for the unclosed terms in the Reynolds averaged equations,
founded on RNG ideas, with (for a finite expansion) a closed form remainder.
RNG methods, by themselves, do not predict the equation coefficients, so that
dynamic LES methods, which do, are an extension of the RNG methodology.
At leading order, this expansion is a step in the derivation of
closures for the quadratic SGS terms, coinciding with the derivation of the
Leonard stress. 
The leading order term of our expansion is related to the Clark model
\cite{ClaFerRey79}. Due to stability problems with the Clark model,
which arise because the SGS terms are not definite,
the exact closures, even in leading order,
are further approximated, most commonly 
by gradient diffusion models in a Smagorinsky
tradition.
The second main contribution in this paper, of a more
speculative nature, consists of observations concerning
the nonuniqueness of the $\mathrm{Re} \rightarrow \infty$ limit,
a phenomenon that is naturally understood within the RNG framework.
We also identify numerical nonuniqueness as an issue for verification
methodology. Here we also make contact with the ILES method, in which explicit
SGS terms are replaced by algorithmic details, resulting in nonunique
(effective, numerically or algorithmically driven)
SGS terms and nonuniqueness of the solution.

We begin with the second result, nonuniqueness,
as the terminology needed for the rest of the
paper is introduced in this manner.
In the RNG line of reasoning,
the Euler equations,
as a limit of the Navier-Stokes equations,
arise as an RNG fixed point.
We have postulated~\cite{LimKamYu12,KamLimYu11} that for the mixing
of $N$ species, 
the high Re limit point is not unique;
rather, it is chosen from an $N+1$ parameter family of limit points,
labeled by points $T \in \mathcal{T}$,
with $\mathcal{T}$ being the $(N+1)$-dimensional space of
transport coefficient ratios for the Navier-Stokes equation,
namely the $N-1$ independent dimensionless
mass diffusivities (Schmidt numbers), the dimensionless
thermal diffusivity (the Prandtl number), and
the dimensionless ratio of the anisotropic and isotropic viscosities.
In the case of multi-temperature thermodynamics,
$\mathcal{T}$ is $2N$ dimensional,
the increase resulting from $N-1$ new Prandtl numbers.
Each sequence $\mathrm{Re} \to \infty$ is accompanied by a sequence
$T(\mathrm{Re}) \in \mathcal{T}$,
which we assume to be convergent: $T(\mathrm{Re}) \to T(\mathrm{Re}=\infty)$.
Dimensional transport coefficients are denoted as $t$. In the limit
$\mathrm{Re} \rightarrow \infty$ as we define it, $t \rightarrow 0$, but other
choices of the limit $\mathrm{Re} \rightarrow \infty$ are possible,
as we illustrate below.
Further comments on nonuniqueness are given in \S\ref{sec:discussion}.

As a simple illustration of the nonuniqueness of the RNG limit, consider
two different physically motivated pictures.
In case~A,
as a model of turbulent combustion,
the thermal diffusion is molecular in nature and enhanced by turbulent mixing.
For this limit, we keep fixed the dimensionless Prandtl number
$\mathrm{Pr} = \nu /\kappa$,
where $\kappa$ is the thermal diffusivity and $\nu$ is the
kinematic anisotropic viscosity, to preserve
the dimensionless aspects of the thermal mixing processes;
the other dimensionless transport coefficients are also fixed in the same
manner. In case~B,
we assume a thermal process unrelated to turbulent mixing (such as radiation),
for which we want to keep $\kappa$ constant so that $\mathrm{Pr} \to 0$
as $\mathrm{Re}\to \infty$. The limiting equations in the two cases~A and~B
do not coincide. For case~A, the limiting equations are the
Euler equations, with all transport coefficients set to zero, and
for case~B, the limiting equations include a thermal diffusion term in the
energy equation.

The full $N+1$ parameter space of RNG limit points is realized by
alternate paths in $\mathcal{T}$, taken as $\mathrm{Re} \to \infty$.
Alternate paths may originate from alternate choices
for the subgrid scale models, or from
alternate physical modeling, leading to alternate settings of the
laminar transport coefficients $t(\mathrm{Re})$, as indicated above
(cases~A and~B).

Related to this physics-level nonuniqueness is a numerical nonuniqueness.
Different numerical algorithms applied to an identical problem yield
apparently converged solutions with significant qualitative
differences~\cite{DimYou04,Mas07,MelRaoKau12a}. Not
only are the SGS terms which select among the $N+1$ parameter family of
solutions commonly omitted or treated implicitly as an aspect of the
algorithm itself, but also numerical truncation errors
contribute at a comparable order
and also participate in the selection of the limit from among the
$N+1$ parameter family of possible solutions. We observe that
nonuniqueness has been demonstrated mathematically for some time~\cite{Sch93a}.

\subsection{Three RNG steps}
\label{sec:RNG-steps}
The RNG has three basic operations:
\begin{enumerate}
\item an integration or coarse-graining step,
which partially solves the equation
in mapping the unknown solution from a fine to a coarser grid;
\item a remapping or rescaling step,
which changes the length scales so that
the unknown length scale of the coarser grid is always fixed;
\item a re-parametrization step,
in which certain solution parameters
(the coefficients of the ``relevant'' terms)
are assigned new, or updated, values to assure
continued agreement with parameters that define the observed solutions.
This agreement is obtained from measurements at the length scale of the
coarse grid.
\end{enumerate}

We propose to label the relevant variables by the $N+1$ parameters
of $\mathcal{T}$; see additional discussion in \S\ref{sec:discussion}.
For each parameter of $\mathcal{T}$, we need an observable, defined
on the coarse grid, to determine that parameter. To establish a bridge to the
LES terminology, we call these observables \emph{models}.

We apply the RNG construction to estimate subgrid scale terms associated with
a grid $\mathcal{M}_n$.
On this grid some terms (namely, the averages of the nonlinear terms in
the Navier-Stokes equations) are unresolved,
in the sense that they fail to be functions only of the averaged primitive
variables. The averaged primitive variables are constant
on each cell of the grid $\mathcal{M}_n$.

The RNG step~1 is
an integration, i.e., a coarse graining step.
This map of solutions induces a map of equations, namely the equations
satisfied by the mesh averages of the solutions. The map of equations
is straightforward as far as linear equation terms are concerned.
(See Secs.~\ref{sec:averaging} and \ref{sec:mixing}.)
But the map leads to unresolved
(unclosed) terms when applied to nonlinear terms in the equation. These
unclosed (as far as $\mathcal{M}_n$ is concerned) terms are identified as
polynomials in first difference operators defined on $\mathcal{M}_{n+1}$.
We iterate on step~1, with step~2 removed. In this way, we start the coarse
graining, in successive RNG iterations,
from $\mathcal{M}_{n+2}, \cdots, \mathcal{M}_{n+k}, \cdots$.
This construction
of unclosed terms for equations on $\mathcal{M}_n$ yields a product of
polynomials in first difference operators on each 
of $\mathcal{M}_{n+1}, \cdots, \mathcal{M}_{n+k}, \cdots$.

In this mapping the SGS terms, the expressions unresolved at
some mesh level $\mathcal{M}_n$ are partially integrated, meaning that their
values on a finer mesh $\mathcal{M}_{n+k}$
are determined, up to expressions unresolved on this finer mesh.
The RNG step~1,
as implemented in this paper,
is to express these terms,
unresolved on $\mathcal{M}_n$,
on successively finer grids $\mathcal{M}_{n+k}$.

In our treatment of step 2, we depart from conventional RNG methodology.
We omit the remap step~2.
For any finite number of steps, the remap is an isomorphism, and so
its presence or absence is moot. However in the limit $k \to \infty$,
there is a difference. The conventionally remapped RNG analysis would
place the LES data at infinitely remote (spatial infinity) wave numbers
to achieve the integration of a truly scale invariant problem.
In the LES framework, we do not wish to eliminate the problem data in this
manner, and so we forgo the remap and the RNG fixed point, while
obtaining a 
RNG based series expansion for the unclosed
$\mathcal{M}_n$ SGS terms. To emphasize this distinction, we refer to the
limit we obtain as an RNG limit point (but not an RNG fixed point).

According to step~3, at each RNG step the
equation parameters (the RNG relevant variables)
should be set equal to the physical parameter
values for that length scale. This must occur at the grid 
level $\mathcal{M}_{n+k}$.

We define turbulent and total dimensional transport parameters, the latter
denoted $t_{\rm total}$.
We take the step~3 requirement to mean that the $N+1$
total (laminar plus turbulent) transport coefficients
$t_{\rm total} \in \mathcal{T}$
should be set at each RNG step. We could as well set the dimensionless
transport coefficients. For this purpose, we follow convention and define
$T_{\rm total} = \nu_{\rm total}/t_{\rm total}$ as the dimensionless
transport parameters.
Sequences with different limit points $T_{\rm total}(\mathrm{Re}=\infty)$
for the dimensionless total transport should generate different
infinite $\mathrm{Re}$ solutions of the Navier-Stokes or Euler equations.

As conventionally applied, for example to quantum field theory, the
relevant variables (particle masses and coupling constants) are predicted
neither by the RNG fixed point nor by the quantum field theory itself.
The mass of the proton, a fundamental particle in strong coupling
quantum field theory, is derived from an extended theory of quarks, in which
the proton is not a fundamental particle but rather a derived object.
More directly comparable is quantum electrodynamics. In principle,
an exact integration of the RNG equation would give the observed
(finite length scale) mass of the electron as a multiple of the
bare (zero length scale) mass, but the bare mass is not known.
We also compare to quantum chromodynamics (QCD), which shares with
turbulence theory the property that the coupling constant (the
turbulent viscosity, for turbulence theory) becomes smaller at short
length scales. This property of QCD is called asymptotic freedom.
In turbulence
theory, a similar exact integration of the RNG defines the
turbulent diffusion coefficient as observed at finite length scales
in terms of the same coefficient as observed at zero length scales. However,
in contrast to quantum electrodynamics, this ``bare'' diffusion coefficient
is generally accepted to be zero, thus is known. Moreover, for turbulent
diffusion,
the RNG integration is dominated by its low order perturbation
term, and thus is effectively determined perturbatively, as with
dynamic SGS models.
Thus the dynamic LES turbulence theory of the infinite $\mathrm{Re}$ limit
goes beyond the quantum field theory RNG in predicting its own coefficients.

\subsection{Models and RNG relevant parameters}
\label{sec:models}
To set parameters according to RNG step~3
at the length scale of $\mathcal{M}_{n+k}$,
we need observables at this same length scale,
and with $N+1$ parameters to be set,
we need to specify $N+1$ observables,
each an observable relative to the mesh $\mathcal{M}_{n+k}$.
In other words, the observables should be
functions of the averaged primitive variables
on the mesh $\mathcal{M}_n$.
These observables constitute the
\emph{model} for the unobserved SGS terms, and RNG step~3 supplies the
(missing) coefficients for the model; i.e., it sets the coefficients
for turbulent transport at each RNG step. In conventional RNG
applications, the model coefficients are set by comparison to
experimental data. Here they are set by theory.
This coefficient specification is equivalent to specifying
a point $T_{\rm total} \in \mathcal{T}$ at each RNG step.

The coefficients for the SGS terms have the form
$\nu_{\, \mathrm{total}}/\mathrm{Sc}_{\,\mathrm{total}}$,
$\nu_{\, \mathrm{total}}/\mathrm{Pr}_{\,\mathrm{total}}$ and a similar
ratio for the isotropic viscosity coefficient.

Dynamic LES is a single expansion step, $k = 1$, of the RNG expansion.
For this step, it contains the same steps 1 and 3
as the RNG and two additional steps:
\begin{enumerate}
\item[4.] Conventionally, we replace the continuous time but space discretized
quantities of the RNG steps 1-3 with a time discretization as well.
Thus the SGS terms are defined as the average over a time step of the
cell face averages introduced below. To define a closed LES algorithm,
we replace these unclosed SGS terms with the corresponding models
multiplied by their
coefficients as defined by step~3 in the numerical solution of the equations
on the grid $\mathcal{M}_n$.
\item[5.] Model coefficients are determined by a theoretical analysis from the
mesh level solutions to the governing equations.
\end{enumerate}

In the RNG formalism, we label expansion terms at each order as
irrelevant, relevant, or nonrenormalizable according to whether they
become smaller, remain constant or grow as the renormalization map is
iterated, and we consider the effects of finer grids as refinements of
the grid $\mathcal{M}_n$. There should be no nonrenormalizable parameters.
We anticipate $N+1$ relevant ones, namely the
$N+1$ coefficients of the models, interpreted as dimensionless total
transport coefficients.

Settings of the relevant parameters follow from the dynamic
SGS analysis~\cite{MoiSquCab91}, a version of which is repeated here.
This method specifies an $\mathrm{Re}$-dependent path
$T_{\rm turbulent}(\mathrm{Re}) \in \mathcal{T}$.
These coefficients are combined with the laminar ones
$T(\mathrm{Re}) \in \mathcal{T}$ through a formula
$t_{\rm total} = t_{\rm turbulent} + t_{\rm laminar}$
and together define $T_{\rm total}(\mathrm{Re}) \in \mathcal{T}$
which uniquely specifies
the LES method.
The dynamic SGS method assumes a functional form for the model, and given that,
derives the coefficients. Variations in this formalism can be based on
alternate formulations for the model. The dynamic SGS method determines
the coefficients of the model directly from
the solutions at grid level $\mathcal{M}_n$ of the
governing equations, and seems
to leave little room for argument
regarding the coefficients, given the model.

\subsection{Outline}
\label{sec:outline}
Our formalism is developed in \S\ref{sec:averaging}
and illustrated for compressible mixing in \S\ref{sec:mixing}.
The expansions for SGS terms are developed in \S\ref{sec:expansions}.
The leading order expansion terms are discussed in \S\ref{sec:leading_order},
with the relation to the Leonard stress given in \S\ref{sec:modeling}.
In \S\ref{sec:numerical},
we explore the possible role of numerical issues
in modifying or selecting the limit point.
A concluding discussion is found in \S\ref{sec:discussion}.

\section{Averaging Procedure}
\label{sec:averaging}

The Reynolds-averaged equations
associated with a system of conservation laws
are obtained through application of a Reynolds averaging operator,
such as ensemble averaging.
Similarly,
the equations that are discretized in an LES method
arise from application of an integral operator,
viz., convolution with a filter.
Both a Reynolds averaging operator and convolution with a filter
commute with spatial and temporal derivatives.

We prefer instead to use the averaging operator defined by
volume integration over the cells of a spatial grid.
Although the time derivative commutes such volume averaging,
spatial derivatives do not.
We compensate by viewing the cell averages
of the solution of the conservation laws
as solving a finite volume approximation to the system.

\subsection{Discretized equations}
\label{sec:discretized}
Let us write the system of conservation laws under consideration as
\begin{equation}
\frac{\partial U}{\partial t}
+ \frac{\partial F_i(U)}{\partial x_i} = 0 \ ,
\label{eq:conservation_law}
\end{equation}
where repeated tensor indices are summed.
Here $U$ denotes the vector of densities of conserved quantities
(e.g., mass, momentum, energy, and species).
To this system we apply
the cell averaging operator
associated with a polygonal finite-volume mesh $\mathcal{M}$.
For a cell $\mathcal{C}$ of $\mathcal{M}$,
let $\overline{a}^{\mathcal{C}}$
denote the volume integral over $\mathcal{C}$ of the quantity $a$
divided by the volume $V(\mathcal{C})$ of $\mathcal{C}$.
By the divergence theorem,
the volume integral over $\mathcal{C}$ of the divergence of the flux $F_i(U)$
equals the surface integral of the outward normal component of the flux
over the boundary $\partial\mathcal{C}$.
This boundary consists of the faces of the cell.
If $f \subset \partial\mathcal{C}$ is a face of $\mathcal{C}$,
let $\overline{b}^f$ denote
the surface integral over $f$ of the quantity $b$
divided by the surface area $A(f)$ of $f$.
Then averaging Eq.~(\ref{eq:conservation_law})
over the cell $\mathcal{C}$ yields
the semi-discrete evolution equation
\begin{equation}
\label{eq:averaged_conservation_law}
\frac{\mathrm{d} \overline{U}^{\mathcal{C}}}{\mathrm{d} t}
+ \sum_{\{f \subset \partial\mathcal{C} : f \mathrm{a~face~of~}\cal{M}_n \}}
\frac{A(f)}{V(\mathcal{C})}
\,n_i^{\mathcal{C},\,f}\,\overline{F_i(U)}^f = 0 \ ,
\end{equation}
where $n_i^{\mathcal{C},\,f}$ is the unit normal to face $f$
pointing outward from $\mathcal{C}$.

The face-averaged flux $\overline{F_i(U)}^f$ is constructed as follows.
First,
we write
\begin{equation}
\label{eq:SGS}
\overline{F_i(U)}^f = F_i(\overline{U}^f) + F_i^{f,\,\mathrm{SGS}}(U) \ .
\end{equation}
This equation defines the SGS flux $F_i^{f,\,\mathrm{SGS}}(U)$.
Explicit formulae for the SGS flux components for compressible mixing
are given in \S\ref{sec:mixing};
an expansion for each SGS term,
which is the central result of the present paper,
is developed in \S\ref{sec:expansions}.
Second,
$F_i(\overline{U}^f)$ is related to
cell-averaged conserved quantities through a numerical scheme,
such as
\begin{equation}
\label{eq:numerical_flux}
n_i^{\mathcal{C},\,f} F_i(\overline{U}^f)
= \mathcal{F}(\overline{U}^{\mathcal{C}},
\overline{U}^{\mathcal{C}'}, n^{\mathcal{C},\,f}\,)
+ \mathcal{E}^{\mathcal{C},\,f}(U) \ ,
\end{equation}
$\mathcal{C}'$ being the cell sharing face $f$ with $\mathcal{C}$.
Here $\mathcal{F}$ is the numerical flux
for a conservative, consistent finite-volume scheme
and $\mathcal{E}^{\mathcal{C},\,f}$ is its truncation error,
discussed in \S\ref{sec:numerical}.

\subsection{Projection operators}
\label{sec:projection}
To be more concrete,
we focus on cubic meshes.
That is,
we take the physical domain to be $\mathcal{D} = [0,1]^D$,
the $D$-dimensional unit cube with periodic boundary conditions,
and we let $\mathcal{M}_n$ be the cubic mesh
with a corner at the origin
that divides $\mathcal{D}$ into equal sized cells,
each with edge lengths $2^{-n}$.
Thus $\mathcal{M}_{n+1}$, $\mathcal{M}_{n+2}$, $\ldots$
are nested refinements of the base mesh $\mathcal{M}_n$,
which remains fixed throughout the discussion.

Let $E_n$ be the operation of averaging over faces of the grid $\mathcal{M}_n$.
Applied to a function $a$ defined on $\mathcal{F}_n$
(the union of the faces of $\mathcal{M}_n$),
this operation yields a piecewise-constant function $E_n\,a$
that is constant over each face,
the constant value for face $f$ being
its average $\overline{a}^f$ over the face.
On the Hilbert space
$\mathcal{H}_n = L^2(\mathcal{F}_n, \mathrm{d}^{D-1}x)$,
the operator $E_n$ is an orthogonal projection onto
the subspace $E_n \mathcal{H}_n$
comprising such piecewise-constant functions.
In fact for face $f$,
\begin{equation}
\overline{a}^f
= \frac{\int_f a\,\mathrm{d}^{D-1}x}{\int_f \mathrm{d}^{D-1}x}
= \frac{\langle 1_f, a \rangle}{\langle 1_f, 1_f \rangle} \ ,
\end{equation}
where $1_f$ is the characteristic function of $f$,
which takes the value $1$ on $f$ and zero elsewhere,
and the brackets $\langle \cdot , \cdot \rangle$
denote the usual inner product on $\mathcal{H}_n$.
Therefore
\begin{equation}
E_n\,a  
= \sum_{\{ f \subset \mathcal{F}_n: f \mathrm{a~face~of~}\cal{M}_n \}}
\overline{a}^f\,1_f
= \sum_{\{ f \subset \mathcal{F}_n: f \mathrm{a~face~of~}\cal{M}_n \}}
\frac{1_f\,\langle 1_f, a \rangle}{\langle 1_f, 1_f \rangle} \ .
\end{equation}
Let $F_n = I - E_n$ denote the complementary projection operator.
When the choice of mesh is unambiguous,
we write $\overline{a}$ in place of the mean $E_n\,a$
and $a' = a - \overline{a}$ in place of the fluctuation $F_n\,a$.

In addition to the simple averaging operator $E_n$,
the mass density-weighted, or Favre, averaging operator
$\widetilde{E}_n$ is useful.
We assume that $\rho$ has sufficient regularity to allow its
restriction to the faces of $\mathcal{M}_n$ and,
as restricted,
to being locally integrable.
On the Hilbert space
$\widetilde{\mathcal{H}}_n = L^2(\mathcal{F}_n, \rho\,\mathrm{d}^{D-1}x)$,
the operator $\widetilde{E}_n$ is likewise an orthogonal projection
onto the subspace $\widetilde{E}_n \widetilde{\mathcal{H}}_n$.
For face $f$
\begin{equation}
\label{eq:rho-a}
\widetilde{a}^f
= \frac{\overline{\rho a}^f}{\overline{\rho}^f}
= \frac{\int_f a \rho\,\mathrm{d}^{D-1}x}{\int_f \rho\,\mathrm{d}^{D-1}x}
= \frac{\langle 1_f, a \rangle_\rho}{\langle 1_f, 1_f \rangle_\rho} \ ,
\end{equation}
where the notation $\langle \cdot , \cdot \rangle_\rho$
stands for the usual inner product on $\widetilde{\mathcal{H}}_n$.
Hence
\begin{equation}
\label{eq:tilde_E_n}
\widetilde{E}_n\,a
= \sum_{f \subset \mathcal{F}_n}
\frac{1_f\,\langle 1_f, a \rangle_\rho}{\langle 1_f, 1_f \rangle_\rho} \ .
\end{equation}
We also let $\widetilde{F}_n = I - \widetilde{E}_n$
and, when unambiguous,
denote the Favre mean $\widetilde{E}_n\,a$ by $\widetilde{a}$
and the Favre fluctuation $\widetilde{F}_n\,a$ by $a'' = a - \widetilde{a}$.

For $k = 1$, $2$, $\ldots$,
the projection operator $\widetilde{E}_{n+k}$ on $\widetilde{\mathcal{H}}_{n+k}
= L^2(\mathcal{F}_{n+k}, \rho\,\mathrm{d}^{D-1}x)$ is defined similarly.
However,
we shall regard $\widetilde{E}_{n+k}$ instead
as the operator on $\widetilde{\mathcal{H}}_n$
that acts in the following way on $a \in \widetilde{\mathcal{H}}_n$:
first extend $a$ to all of the faces of $\mathcal{M}_{n+k}$,
setting it to zero except on the faces of $\mathcal{M}_n$;
next apply $\widetilde{E}_{n+k}$;
and finally restrict the result to the faces of $\mathcal{M}_n$,
obtaining $\widetilde{E}_{n+k}\,a \in \widetilde{\mathcal{H}}_n$.
The operator $\widetilde{E}_{n+k}$ so defined is a projection operator
on $\widetilde{\mathcal{H}}_n$.
Thus $\widetilde{E}_{n+k} \widetilde{\mathcal{H}}_n$ consists of
square-integrable functions defined on $\mathcal{F}_n$
that are constant on each face of $\mathcal{M}_{n+k}$
contained within a face of $\mathcal{M}_n$.

We note an operator identity that will be useful:
\begin{equation}
\widetilde{E}_n = \widetilde{E}_n \widetilde{E}_{n+1} \ .
\label{eq:E_identity}
\end{equation}
Also,
because $\widetilde{E}_n$ is a projection operator
given by Eq.~(\ref{eq:tilde_E_n}),
\begin{equation}
\langle 1_f, a\,\widetilde{E_n}\,b \rangle_\rho
= \langle \widetilde{E_n}(1_f\,a), \widetilde{E_n}\,b \rangle_\rho
= \langle 1_f, (\widetilde{E_n}\,a)\,\widetilde{E_n}\,b \rangle_\rho \ ,
\end{equation}
so that
\begin{equation}
\label{eq:flip}
\widetilde{E}_n\left[a\,\widetilde{E}_n\,b\right]
= \widetilde{E}_n\left[(\widetilde{E}_n\,a)\,\widetilde{E}_n\,b\right] \ .
\end{equation}

\section{Averaged Equations for Compressible Mixing}
\label{sec:mixing}

To illustrate the SGS flux defined by Eq.~(\ref{eq:SGS}),
we describe it in detail for the system of conservation laws
governing the compressible mixing of two gases.

The state of an ideal mixture of two polytropic gases
is characterized by the field variables
$\rho$, $v_i$, $T$ and $\psi$,
which denote the mass density, particle velocity vector,
temperature and mass fraction, respectively.
The mixture has pressure $p = \rho R T$,
where $R = \psi R_1 + (1 - \psi) R_2$,
and specific internal energy $e = c_v T$,
where $c_v = \psi\,c_{v,1} + (1 - \psi)\,c_{v,2}$.
Here $R_\alpha = N_A k_B / M_\alpha$ and $c_{v,\alpha}$ are constants
for $\alpha = 1$, $2$.
The specific total energy is denoted
$E = \tfrac{1}{2} v_\ell v_\ell + e$.
The specific enthalpy of the mixture is
$h = e + p/\rho = c_p T$ with $c_p = c_v + R$,
and the individual specific enthalpies for the two gases are
$h_\alpha = c_{p,\alpha} T$,
where $c_{p,\alpha} = c_{v_,\alpha} + R_\alpha$.

The laminar transport coefficients for momentum, heat, and mass
are the kinematic anisotropic and isotropic viscosities,
$\nu$ and $\nu_{\mathrm{i}}$,
thermal diffusivity $\kappa = \nu/\mathrm{Pr}$
and mass diffusivity $D = \nu/\mathrm{Sc}$,
where $\mathrm{Pr}$ and $\mathrm{Sc}$ denote
the Prandtl and Schmidt parameters.
The viscous stress tensor,
heat flux vector,
and diffusive mass flux vector are
\begin{align}
\label{eq:viscous_stress}
\sigma^{\mathrm{v}}_{ij} &=
\rho \nu_{\mathrm{i}}\,\frac{\partial v_\ell}{\partial x_\ell}\,\delta_{ij}
+ \rho \nu \left ( \frac{\partial v_i}{\partial x_j}
+ \frac{\partial v_j}{\partial x_i}
- \frac{2}{3}\frac{\partial v_\ell}{\partial x_\ell}\,\delta_{ij} \right)\ , \\
\label{eq:heat_flux}
q_i &= - \rho \kappa c_p\,\frac{\partial T}{\partial x_i} \ , \\
\label{eq:diffusive_mass_flux}
j_i &= - \rho D\,\frac{\partial \psi}{\partial x_i} \ .
\end{align}
For simplicity,
$\rho \nu$, $\rho \nu_{\mathrm{i}}$, $\rho \kappa$ and $\rho D$
(hence $\nu/\nu_{\mathrm{i}}$, $\mathrm{Pr}$ and $\mathrm{Sc}$)
are assumed to be constant.

The cell-averaged laws of conservation of mass, momentum, energy and species,
which govern the averaged field variables $\overline{\rho}$,
$\widetilde{v_i}$, $\widetilde{T}$ and $\widetilde{\psi}$,
are
\begin{align}
\label{eq:gas-0}
&\frac{\partial\,\overline{\rho}}{\partial t}
+ \frac{\partial\,\overline{\rho}\,\widetilde{v_i}}{\partial x_i} = 0 \ , \\
\label{eq:gas-1}
&\frac{\partial\,\overline{\rho}\,\widetilde{v_j}}{\partial t}
+ \frac{\partial(\overline{\rho}\,\widetilde{v_i}\,\widetilde{v_j}
+ \overline{p}\,\delta_{ij})}{\partial x_i}
= \frac{\partial \overline{\sigma^{\mathrm{v}}_{ij}}}{\partial x_i}
- \frac{\partial \tau_{ij}}{\partial x_i} \ , \\
\label{eq:gas-2}
&\frac{\partial\,\overline{\rho} \widetilde E}{\partial t}
+ \frac{\partial(\overline{\rho} \widetilde E + \overline{p})
\,\widetilde{v_i}}{\partial x_i}
= \frac{\partial \overline{\sigma^{\mathrm{v}}_{ij}}\,\widetilde{v_j}}
{\partial x_i} - \frac{\partial \overline{q_i}}{\partial x_i}
- \frac{\partial (\widetilde{h_1} - \widetilde{h_2})\,\overline{j_i}}
{\partial x_i} \nonumber \\
&\qquad\qquad\qquad
- \frac{\partial \tau_{ij} \widetilde{v_j}}{\partial x_i}
- \frac{\partial q_i^{(h)}}{\partial x_i}
- \frac{\partial q_i^{(h \psi)}}{\partial x_i}
- \frac{\partial q_i^{(k)}}{\partial x_i}
- \frac{\partial q_i^{(\mathrm{v})}}{\partial x_i} \ , \\
\label{eq:gas-3}
&\frac{\partial\,\overline{\rho}\,\widetilde{\psi}}{\partial t}
+ \frac{\partial\,\overline{\rho}\,\widetilde{\psi}
\,\widetilde{v_i}}{\partial x_i}
= - \frac{\partial \overline{j_i}}{\partial x_i}
- \frac{\partial q_i^{(\psi)}}{\partial x_i} \ .
\end{align}
For clarity,
we have made the formal replacement (which is actually an identity
in the sense of distribution derivatives)
\begin{equation}
\sum_{\{ f \subset \partial\mathcal{C}: f \mathrm{is~a~face~of}{\cal{M}_n}\}}
\frac{A(f)}{V(\mathcal{C})}
\,n_i^{\mathcal{C},\,f} \mapsto \frac{\partial}{\partial x_i}
\end{equation}
in Eq.~(\ref{eq:averaged_conservation_law})
and omitted the $\mathcal{C}$ and $f$ indications
on the averaging operators.
(An average of a field variable appearing within a time derivative
is a cell average,
whereas within a spatial derivative it is a face average.)

Appearing in Eqs.~(\ref{eq:gas-0})--(\ref{eq:gas-3})
are the SGS flux components $\tau_{ij}$, $q_i^{(h)}$, $q_i^{(h \psi)}$,
$q_i^{(k)}$, $q_i^{(\mathrm{v})}$ and $q_i^{(\psi)}$,
which are defined by
\begin{align}
\label{eq:tau}
\tau_{ij} &= \overline{\rho} \left(\widetilde{v_i\,v_j}
- \widetilde{v_i}\,\widetilde{v_j}\right)
= \overline{\rho}\,\widetilde{v_i''\,v_j''} \ , \\
\label{eq:qh}
q_i^{(h)} &= \overline{\rho} \left(\widetilde{h\,v_i}
- \widetilde{h}\,\widetilde{v_i}\right)
= \overline{\rho}\,\widetilde{c_p}\,\widetilde{T''\,v_i''}
+ \overline{\rho}\,(c_{p,1} - c_{p,2})
\left(\widetilde{T}\,\widetilde{\psi''\,v_i''}
+ \widetilde{\psi''\,T''\,v_i''}\right) \ , \\
\label{eq:qh_psi}
q_i^{(h \psi)} &= \overline{(h_1 - h_2)\,j_i}
- (\widetilde{h_1} - \widetilde{h_2})\,\overline{j_i}
= (c_{p,1} - c_{p,2})\left(\overline{T''}\,\,\overline{j_i}
+ \overline{T' j_i'}\right) \ , \\
\label{eq:qT}
q_i^{(k)} &= \tfrac{1}{2} \overline{\rho} \left(\widetilde{v_\ell\,v_\ell\,v_i}
- 2\,\widetilde{v_\ell}\,\widetilde{v_\ell\,v_i}
- \widetilde{v_\ell\,v_\ell}\,\widetilde{v_i}
+ 2\,\widetilde{v_\ell}\,\widetilde{v_\ell}\,\widetilde{v_i}\right)
= \tfrac{1}{2} \overline{\rho}\,\widetilde{v_\ell''\,v_\ell''\,v_i''} \ , \\
\label{eq:qv}
q_i^{(\mathrm{v})} &= - \left(\overline{\sigma^{\mathrm{v}}_{ij}\,v_j}
- \overline{\sigma^{\mathrm{v}}_{ij}}\,\widetilde{v_j}\right)
= - \overline{\sigma^{\mathrm{v}}_{ij}}\,\,\overline{v_j''}
- \overline{(\sigma^{\mathrm{v}}_{ij})'\,v_j'} \ , \\
\label{eq:qpsi}
q_i^{(\psi)} &= \overline{\rho} \left(\widetilde{\psi\,v_i}
- \widetilde{\psi}\,\widetilde{v_i}\right)
= \overline{\rho}\,\widetilde{\psi''\,v_i''} \ .
\end{align}
Here we have taken advantage of the properties
$\overline{a'} = 0$ and $\widetilde{a''} = 0$ of cell and face averaging
(but not convolution with a filter).

A related SGS quantity,
the fluctuation kinetic energy $k$
defined by $\overline{\rho}\,k = \tfrac{1}{2} \tau_{\ell\ell}
= \tfrac{1}{2} \overline{\rho}\,\widetilde{v_\ell''\,v_\ell''}$,
arises in the formula $\widetilde{E}
= \tfrac{1}{2} \widetilde{v_\ell}\,\widetilde{v_\ell} + \widetilde{e} + k$.
The Favre-averaged specific internal energy $\widetilde{e}$
is viewed as $\widetilde{c_v} \widetilde{T}$
plus the SGS term $\widetilde{c_v T} - \widetilde{c_v} \widetilde{T}
= (c_{v,1} - c_{v,2})\,\widetilde{\psi''\,T''}$;
and the averaged pressure $\overline{p} = \overline{\rho} \widetilde{R\,T}$
occurring in the conservation laws is treated similarly.
Also,
in $\overline{j_i} = - \rho D\,\partial \overline{\psi}/\partial x_i$,
the variable $\overline{\psi}$ should be replaced by $\widetilde{\psi}$
plus the SGS term $\overline{\psi''}$;
similarly,
the SGS terms $\overline{v_\ell''}$ and $\overline{T''}$
enter into $\overline{\sigma^{\mathrm{v}}_{ij}}$
and $\overline{q_i}$,
respectively.
Notice that a quantity such as $\overline{\psi''}$ is equivalently written as
$\overline{\rho}\,\widetilde{(\rho^{-1})''\,\psi''}$.

Thus we see that an SGS term for the compressible mixing of two gases,
under the stated modeling assumptions,
involves either a covariance $\cov(a,b) = \overline{a'\,b'}$,
a Favre covariance $\widetilde{\cov}(a, b) = \widetilde{a''\,b''}$,
or a third-order Favre cumulant
$\widetilde{\cum}(a, b, c) = \widetilde{a''\,b''\,c''}$,
where $a$, $b$ and $c$ are each one of $\rho^{-1}$, $v_i$, $T$ or $\psi$
or their spatial derivatives.

\section{Expansions for SGS Terms}
\label{sec:expansions}

The flux appearing in the discretized conservation
laws~(\ref{eq:averaged_conservation_law}) is decomposed,
via Eq.~(\ref{eq:SGS}),
into a function of $\overline{U} = E_n\,U$
plus the SGS flux.
A function of $E_n\,U$ is said to be resolved on $\mathcal{M}_n$.
In this section we develop expansions for SGS terms
involving quantities that are resolved on successively finer grids.

\subsection{Leonard SGS term}
\label{sec:leonard}
A familiar SGS term is the Reynolds stress tensor $\tau^{(n)}_{ij}$,
where
\begin{equation}
\tau^{(n)}_{ij}
= \overline{\rho\,v_i\,v_j}
- \overline{\rho\,v_i}\ \overline{\rho\,v_j}/\overline{\rho} \ .
\end{equation}
The superscript $n$ indicates that the averaging occurs
on the grid $\mathcal{M}_n$.
Consider the corresponding tensor
$\tau^{(n-1)}_{ij} = \widehat{\rho\,v_i\,v_j}
- \widehat{\rho\,v_i}\ \widehat{\rho\,v_j}/\widehat{\rho}$
on the once-coarsened grid $\mathcal{M}_{n-1}$,
where the caret denotes averaging for $\mathcal{M}_{n-1}$,
and compare it with $\tau^{(n)}_{ij}$,
as averaged onto $\mathcal{M}_{n-1}$.
The Germano identity~\cite{Ger90} is that the difference
$\tau^{(n-1)}_{ij} - \widehat{\tau^{(n)}_{ij}}$
between these two $\mathcal{M}_n$-unresolved SGS terms
reduces to the Leonard stress tensor~\cite{Leo74,MoiSquCab91,Ma06}
\begin{equation}
\mathcal{L}^{(n-1)}_{ij}
= \widehat{\overline{\rho}\,\tilde v_i\,\tilde v_j}
-  \widehat{\overline{\rho}\,\tilde v_i}\ \widehat{\overline{\rho}\,\tilde v_j}
/ \widehat{\overline{\rho}} \ ,
\end{equation}
which is resolved on $\mathcal{M}_n$.
The reduction occurs because of the cancellation
of the two terms,
$\widehat{\rho\,v_i\,v_j}$ and $\widehat{\overline{\rho\,v_i\,v_j}}$,
that are not $\mathcal{M}_n$-resolved,

\subsection{Generalized Leonard SGS term}
\label{sec:generalized_leonard}
More generally,
consider an SGS term in the form of a Favre covariance
\begin{equation}
\label{eq:covariance}
\widetilde{\cov}_n(a, b)
= \widetilde{a''\,b''}
= \widetilde{a\,b''}
= \widetilde{E}_n\left[a\,\widetilde{F}_n\,b\right]
\end{equation}
of the quantities $a$ and $b$
with respect to the mesh $\mathcal{M}_n$.
(The ordinary covariance $\cov_n(a, b) = \overline{a'\,b'}$
is included as a special case.)
By analogy,
we define the corresponding Leonard SGS term to be
\begin{equation}
\label{eq:L_tilde}
\widetilde{\mathcal{L}}_n(a, b)
= \widetilde{\cov}_n(a, b)
- \widetilde{E}_n\,\widetilde{\cov}_{n+1}(a, b) \ .
\end{equation}
As we now demonstrate,
this quantity is $\mathcal{M}_{n+1}$-resolved
provided that $\rho\,a$ and $\rho\,b$ are components of $U$.

Using the identity~(\ref{eq:covariance}) on levels $n$ and $n+1$,
along with the operator identity~(\ref{eq:E_identity})
and the definitions $\widetilde{F}_n = I - \widetilde{E}_n$
and $\widetilde{F}_{n+1} = I - \widetilde{E}_{n+1}$,
we see that
\begin{equation}
\begin{split}
\widetilde{\mathcal{L}}_n(a, b)
&= \widetilde{E}_n \left[a\,\widetilde{F}_n\,b \right]
- \widetilde{E}_n \widetilde{E}_{n+1}\left[a\,\widetilde{F}_{n+1}\,b\right] \\
&= \widetilde{E}_n \left[a\,(\widetilde{E}_{n+1}-\widetilde{E}_n)\,b\right] \ .
\end{split}
\end{equation}
By identity~(\ref{eq:E_identity}) again
along with Eq.~(\ref{eq:flip}) at levels $n$ and $n+1$,
\begin{equation}
\widetilde{\mathcal{L}}_n(a, b)
= \widetilde{E}_n
\left[(\widetilde{E}_{n+1}\,a)\,\widetilde{E}_{n+1}\,b\right]
- \widetilde{E}_n
\left[(\widetilde{E}_n\,a)\,\widetilde{E}_n\,b\right] \ .
\end{equation}
Finally,
according to identity~(\ref{eq:E_identity}),
an $\mathcal{M}_n$-resolved quantity is also $\mathcal{M}_{n+1}$-resolved.
Hence $\widetilde{\mathcal{L}}_n(a, b)$ is $\mathcal{M}_{n+1}$-resolved.

\subsection{Expansion for a covariance}
\label{sec:covariance}
Next we develop an expansion for the Favre covariance
$\widetilde{\cov}_n(a, b) = \widetilde{a''\,b''}$.
By definition of the generalized Leonard SGS term,
\begin{align}
\label{eq:sgs}
\widetilde{\cov}_n(a, b)
= \widetilde{\mathcal{L}}_n(a, b)
+ \widetilde{E}_n\,\widetilde{\cov}_{n+1}(a, b) \ .
\end{align}
By induction on $n$,
we see that
\begin{equation}
\label{eq:expansion}
\widetilde{\cov}_n(a, b)
= \widetilde{E}_n\,\sum_{j = 0}^J \widetilde{\mathcal{L}}_{n+j}(a, b)
+ \widetilde{E}_n\,\widetilde{\cov}_{n+J+1}(a, b) \ .
\end{equation}
The sum involving Leonard terms,
which we denote by $\widetilde{\cov}_{n,J}(a,b)$,
is resolved on the grid $\mathcal{M}_{n+J+1}$.
The remainder is denoted by $R_{n,J}(a,b)$.

\subsection{Expansion for a third-order cumulant}
\label{sec:cumulant}
Some of the SGS terms in system~(\ref{eq:gas-0})--(\ref{eq:gas-3})
involve third-order cumulants,
such as $\widetilde{\psi''\,T''\,v_i''}$
and $\widetilde{v_\ell''\,v_\ell''\,v_i''}$.
The presence of fluctuations indicates that
these terms are not $\mathcal{M}_n$-resolved;
and because they are not covariances,
they are not handled by the methods of \S\ref{sec:covariance}.
Our procedure is to expand each fluctuation factor
as an average plus a fluctuation relative to $\mathcal{M}_{n+1}$,
and after organizing and simplifying the result,
we iterate and generate the expansion with remainder to all orders.

The expansion step is to replace each of the $\widetilde{F}_n$ operators
in the general third-order cumulant,
\begin{equation}
\widetilde{\cum}_n(a, b, c)
= \widetilde{a''\,b''\,c''}
= \widetilde{E}_n\left[(\widetilde{F}_n a)\,(\widetilde{F}_n b)
\,(\widetilde{F}_n c)\right] \ ,
\end{equation}
using the identity
\begin{equation}
\widetilde{F}_n
= \left(\widetilde{E}_{n+1} - \widetilde{E}_n\right)
+ \widetilde{F}_{n+1} \ .
\end{equation}
Let the quantity in parentheses be denoted by $\widetilde{F}_{n,1}$.
The expansion generates eight terms.
The term
\begin{equation}
\widetilde{\cum}_{n,1}(a, b, c)
= \widetilde{E}_n\left[(\widetilde{F}_{n,1}\,a)
\,(\widetilde{F}_{n,1}\,b)\,(\widetilde{F}_{n,1}\,c)\right]
\end{equation}
from which $\widetilde{F}_{n+1}$ is absent
is the leading order of the expansion.
The seven terms with one, two or three
$\widetilde{F}_{n+1}$ operators
constitute the remainder term,
$\widetilde{R}_{n,1}(a, b, c)$.
In fact,
the terms in $\widetilde{R}_{n,1}(a, b, c)$
with one $\widetilde{F}_{n+1}$ are zero as they each involve
a fluctuation averaged against the product of two constant states;
thus there are four non-zero remainder terms.

To continue this expansion of $\widetilde{a''\,b''\,c''}$,
we employ the operator identity
\begin{equation}
\label{eq:expansion_F}
\widetilde{F}_n
= \left(\widetilde{E}_{n+J+1} - \widetilde{E}_n\right)
+ \widetilde{F}_{n+J+1} \ .
\end{equation}
With the quantity in parentheses in the preceding equation
denoted by $\widetilde{F}_{n,J}$,
\begin{equation}
\label{eq:expansion_3}
\widetilde{\cum}_n(a, b, c)
= \widetilde{E}_n\left[(\widetilde{F}_{n,J}\,a)\,(\widetilde{F}_{n,J}\,b)
\,(\widetilde{F}_{n,J}\,c)\right] + \widetilde{R}_{n,J}(a,b,c) \ ,
\end{equation}
where the remainder $\widetilde{R}_{n,J}(a,b,c)$ comprises
the four nonzero terms
involving the operator $\widetilde{F}_{n+J+1}$.
The first term in Eq.~(\ref{eq:expansion_3}),
which we denote by $\widetilde{\cum}_{n,J}(a, b, c)$,
is resolved on the grid ${\mathcal M}_{n+J+1}$.

\section{Modeling}
\label{sec:modeling}

The discretized system of conservation laws,
Eqs.~(\ref{eq:averaged_conservation_law})--(\ref{eq:numerical_flux}),
involves the SGS flux,
defined by Eq.~(\ref{eq:SGS}),
which depends in an essential way on the solution $U$,
not solely on the face average $\overline{U} = E_n\,U$.
To close the governing system of equations,
each SGS term must be related to the face average of the solution
through a closure relation.
In dynamic SGS modeling,
the closure relation is determined with the aid of the Leonard SGS term.

\subsection{Dynamic SGS modeling}
\label{sec:dynamic}
We assume that the replacement for the SGS term
for the grid level $\mathcal{M}_n$ takes the form $c_n\,M_n$
for some coefficient sequence $c_n$.
Here $M_n$,
is called the model for the particular SGS term.
It is required to be 
a function of the cell averages of the primitive quantities in the
dynamic equations, so that the expression for $M_n$ ``closes''.
Furthermore,
we make the asymptotic assumption that $c_n$ converges as $n \to \infty$,
so that $c_n$ is approximately independent of $n$ when $n$ is large.
The limit,
denoted $c$,
is the turbulent transport coefficient.
For instance,
when the SGS term is $\widetilde{\cov}_n(a,b)$,
the closure, or modeling, relation is
\begin{equation}
\label{eq:closure}
\widetilde{\cov}_n(a,b) \approx c\,M_n \ .
\end{equation}
If $\widetilde{\cov}_n(a,b)$ is a tensor quantity,
so is the corresponding model $M_n$,
and distinct coefficients are used for
the deviatoric and spherical parts of this relationship.

Determination of the turbulent transport coefficient $c$
(LES step~5 in \S\ref{sec:models}) proceeds as follows.
Because the same coefficient $c$ relates the SGS term to the model
at both grid levels $\mathcal{M}_{n-1}$ and $\mathcal{M}_n$,
the difference $M_{n-1} - \widetilde{E}_{n-1}\,M_n$
is a model for the Leonard SGS term
$\widetilde{\mathcal{L}}_{n-1}(a,b)$.
We therefore compute $c$ by requiring that
\begin{equation}
\label{eq:coeff}
\widetilde{\mathcal{L}}_{n-1}(a,b)
\approx c\,\left(M_{n-1} - \widetilde{E}_{n-1}\,M_n\right) \ .
\end{equation}
All expressions in Eq.~(\ref{eq:coeff}), other than $c$,
are $\mathcal{M}_n$-resolved. Thus, using (\ref{eq:coeff}),
$c$ is also.
For tensor SGS terms,
$c$ is a scalar and~(\ref{eq:coeff}) is interpreted
in the sense of least squares.

The cancellation of terms in $\widetilde{\mathcal{L}}_{n-1}(a,b)$ is exact,
whereas the assumption that $\widetilde{\cov}_n(a,b)$
is proportional to the model $M_n$ is an ansatz or approximation.
The choice of model remains open for experimentation and improvement.
Once the model has been specified,
Eq.~(\ref{eq:coeff}) determines the coefficient $c$ dynamically
from the solution.

In keeping with the discussion in \S\ref{sec:models}
regarding choice of renormalization length scales,
the turbulent transport coefficients of all SGS terms,
excepting the anisotropic viscous term,
are taken to be proportional to the anisotropic turbulent viscosity,
with a dimensionless coefficient of proportionality.

\subsection{Third-order cumulants and higher order expansion terms}
\label{sec:cubic}
We believe the closed form expansion terms derived here for the
unclosed SGS terms will aid in future efforts to improve modeling
of closure terms.

\old{
\section{Convergence of the Expansion}
\label{sec:convergence}

Here we assume 
K41 \cite{Kol41} in the sense that we consider a velocity 
Field $v \in H_\delta$, for a Sobolev space $H_\delta$ of positive index 
$\delta$, whether $v$ is the solution of the Euler equation or not.

Theorem. Assume $v \in H_\delta$, $\delta > 0$ with periodic boundary
conditions for the unit cube. Then the constant density Reynolds stress
defined by (\ref{eq:sgs}) is a continuous functional on $H_\delta$. The series
(\ref{eq:expansion}) with remainder is convergent as $j \rightarrow \infty$
in $H_\delta$
and $\lim_{n \rightarrow \infty} R_n = 0$ in the sense of pointwise convergence.

Proof.
The continuity of $v$ is established in \cite{CheGli12}, which allows
the trace, restriction to a face of a cell in ${\mathcal M}_n$ as a continuous
functional on $H_{\delta}$.
Let us make this restriction, so that $R_n$ is defined by
(\ref{eq:sgs}), with error term $R_{n,J}$ defined in terms of the tangential
direction projection operator $F_{n+j}$. Fourier analysis of $F_{n+j}$
in the tangential directions
has it as a multiplication operator in Fourier space, and multiplying
by zero for all small wave numbers. It follows that $F_{n+j}$ converges
strongly to zero as an $L_2$ operator on a face, in the tangential directions,
as $j \rightarrow \infty$ and pointwise convergence as $n \rightarrow \infty$
follows.
\old{
The SGS term has the form
\begin{equation}
\label{eq:SGS-z}
\left ( \overline{ab}^{f(z)} - 
\overline{a}^{f(z)}~\overline{b}^{f(z)} \right ) |^{z=z_{i+1}}_{z=z_i} \ .
\end{equation}
Here $a,b$ are components of the velocity field $v_j$, and the face $f(z)$,
normal to the direction $z$, is $z$ direction translated, and thus is
considered as a function of $z$. 
In the Sobolev space of $z$ direction negative
index, the evaluations at $z = z_i$ and $z = z_{i+1}$
are continuous functionals.
Our proof applies to a fixed 
vector values function
$v$, which is assumed to be in $L_2$ on restriction to the faces in question.

According to 
(\ref{eq:expansion}), we estimate the associated term from
\begin{equation}
R_{n,J}=E_n \cov_{n+J+1}(a,b) = E_n \left [ a F_{n+J+1} b \right ] \ ,
\end{equation}
which be written as
\begin{equation}
\label{eq:R1}
\int a(x,y,z,t)(F_{n+J+1;x,y}b)(x,y,z,t)dxdydt
\end{equation}
where $F_{n+J+1;x,y}$ is the projector for the face $f(z)$
onto functions orthogonal, in the sense of $x,y$ integration,
to functions constant on ${\mathcal M}_{n+J+1}$.

This projector
maps $b$ onto a function which has mean value zero on each 
$\mathcal{M}_{n+J+1}$ subface of $f(z)$. In Fourier space for the
$x,y$ directions, this projector sets to zero all wave number coefficients
for wave numbers $k_x$ and $k_y$ for which both $|k_x|$ and $|k_y|$
are less than $n+J+1$. According to our assumed energy bounds, realized in a 
Sobolev space of negative index, this quantity has a convergent
estimate as $J \rightarrow \infty$ for fixed $n$ and also as 
$n \rightarrow \infty$ uniformly in $J$. Thus the series expansion
converges as $J \rightarrow \infty$ for fixed $n$ and as $n \rightarrow \infty$
uniformly in $J$.
}

The proof applies to any finite energy ($L_2$) vector field, whether a solution
of the Euler equation or not, and can be interpreted as stating that the
turbulent viscosity, when measured at infinitesimal length scales
(the ``bare'' turbulent viscosity), is zero.
}

\section{The Leading Order RNG Expansion}
\label{sec:leading_order}

To leading order, we
consider a single RNG step.
With the current grid level denoted $\mathcal{M} = \mathcal{M}_n$, we also
consider the once refined grid $\mathcal{M}_{n+1}$, for which each cell has
been refined once in each mesh direction. On
the cell faces of $\mathcal{M}_n$, we consider functions which are
piecewise constant on cell faces of $\mathcal{M}_{n+1}$.
In all cases, the leading order contribution to the unclosed
SGS term is a product of two or more
differences of primitive variables multiplied by an expression
depending on $\rho$ and
perhaps other variables which are not differenced. The differences occur
in one or both of the directions tangential to the face of the grid cell.
See (\ref{eq:quad-explicit},~\ref{eq:FF-evaluate3}).

Detailed evaluation of these terms to offer possible improvement
on SGS modeling will be the subject of a future publication.

\section{Selection of SGS Terms}
\label{sec:numerical}

This section brings together the major themes of this paper:
Nonuniqueness for solutions of a dynamic evolution problem, whether to use or
omit SGS terms, their selection, if to be used, the role of the RNG 
as framework for understanding the selection of SGS terms, and
the verification and validation (V\&V) of a specific recipe (dynamic SGS)
for the selection of SGS terms.

Without question, turbulence and turbulent mixing are important problems
for scientific computing, which plays a central role in modern
engineering design. Validation of these simulations (comparison to
experiment) is an essential part
of the scientific method. With nonunique and
algorithmically dependent solutions, validation is hardly
possible and instead, robust engineering design relies on calibration,
which depends intrinsically 
on a sufficient range of experiments in a neighborhood of a design point.

We see a tight linkage between SGS terms and nonuniqueness. Solutions are
observed to vary as a consequence of variation of the
SGS terms. Omission of these terms, and their replacement by
algorithmic artifacts also leads to a variation in the solution, i.e.,
nonuniqueness. Thus we find that specification of the SGS terms 
(in combination with some control over the numerical artifacts, primarily
numerical diffusion)
removes the nonuniqueness; nonuniqueness thus has its ultimate
origin in underspecification of fluid transport. In the present case,
with molecular transport added where convenient or necessary 
according to the laws of physics, it is the
underspecified turbulent transport which gives rise to nonuniqueness of
solutions, and the SGS terms, which specify subgrid turbulent transport,
restore uniqueness. The RNG provides a framework for understanding this
range of issues. Finally, having understood the problem, the dynamic
selection of SGS terms provides a unique recipe for their selection.
It then remains to show that this choice 
(dynamic SGS terms and control over numerical mass diffusion)
satisfies the standards of V\&V.

\subsection{Nonuniqueness of Solutions}
\label{sec:nonuniqueness}

Nonuniqueness is a strictly mathematical
verification issue, which we address with equations specified at the
continuum level of physics, ignoring considerations of 
kinetic theory
and atomic length scales. 
We distinguish between macro and micro nonuniqueness. The distinction has to 
do with the observable used to measure nonuniqueness and its length
scale in relation to $\Delta x$, in the limit $\Delta x \to 0$.
An observable $\mathcal{O}$ (such as the RT 
instability growth rate $\alpha$) is called \emph{macro}
if its associated length scale $l_\mathcal{O}$ satisfies
$l_\mathcal{O} \gg \Delta x$ and it is called \emph{micro}
(such as atomic mixing properties of the flow)
if $l_\mathcal{O} \le \Delta x$. We distinguish
between mathematical nonuniqueness of the infinite Reynolds number Euler 
equations and the nonuniqueness of apparently
converged LES simulations. We call the latter apparent LES nonuniqueness.

\subsubsection{Examples of Nonuniqueness for Equations of Evolution}
\label{sec:history}

Mathematical nonuniqueness for solutions of the Euler equations
is known~\cite{Sch93a,DeLSze09,DeLSze10}.
An understanding of the nature and origin of this nonuniqueness
is presented in the survey~\cite{DeLSze12}.
It is an open problem to determine
the relevance to physics, if any, of the mathematical nonuniqueness theories.
No features of these non-unique solutions are known which disqualify
them from the point of view of either mathematics or physics.
The strong convergence~\cite{CheGli12} of Navier-Stokes solutions
to the Euler limit, even if extended
conceptually to compressible flows, does not address uniqueness,
and cannot, in so far as nonuniqueness is known mathematically.

Turning to physically relevant nonuniqueness,
turbulent mixing 
is not the first time in which nonuniqueness has been an essential
feature of a mathematical model of time dependent physical phenomena.
Shock refraction problems,
which describe self similar time dependent
solutions, sometimes have multiple solutions.
We note the nonuniqueness of the flame speed when analyzed at the
level of the Euler equations. This nonuniqueness is removed by consideration
of the relation between viscosity and thermal diffusion, in dimensionless
terms the Prandtl number. Detonation waves exhibit multiple
solutions, denote as weak or strong detonations~\cite{CouFri67}.
Often some dissipative mechanism
serves to select the weak detonation. Equations suggested by three phase
flow models for petroleum reservoirs also show nonuniqueness for initial
value problems associated with wave interactions
(Riemann problems)~\cite{AzeMar95,AzeMarPlo96}.
Within the study of turbulence, the sensitivity of 
solutions to turbulence models is ubiquitous. We have thus noted noted
precedents for nonuniqueness of solutions to time dependent problems when
analyzed at the Euler level. This ambiguity is often removed when the
modeled at the Navier-Stokes level. However, for turbulence modeling, the
analogous resolution of ambiguity requires specifying the turbulent transport,
exactly the quantity which introduces the ambiguity.
Nonuniqueness of apparently converged LES
is observed in practice and is an issue
that solution verification methodology has yet to address
\cite{MelRaoKau12a}.

\subsubsection{Theoretically Inferred Nonuniquenes}
\label{sec:theory}

The conceptual analysis of cases~A and~B
of \S\ref{sec:rng-les} suggests nonuniqueness
para\-metrized by points of the space $\mathcal{T}$ of dimensionless total 
(molecular plus turbulent) transport.

\subsubsection{Numerically Observed Nonuniqueness}
\label{sec:numerically_observed}

A wide range of simulation results have been proposed as solutions for
an identical high $\mathrm{Re}$ turbulent mixing problem~\cite{DimYou04}.
Code comparison~\cite{Mas07} of apparently converged solutions
showed differences in thermal mixing properties.
Additionally, systematic variation of the SGS coefficient has been observed
to change the atomic mixing properties of
nominally converged solutions~\cite{MelRaoKau12a}. On this basis, we
consider here the extent to which the selection of
the high $\mathrm{Re}$
limit might be influenced by numerical considerations.
Consistent with this point of view are comments from~\cite{HonMoi05}:
``Results using the MILES approach for LES are found to strongly depend
on scheme parameters and grid size. Also, physical variables cannot be
simultaneously predicted.''
See also~\cite{SilGraMet93,GarMosSag99,CamHonMoi02}.
In Table~\ref{table:RT-alpha},
we summarize the observed numerical and experimental variation
in efforts to determine the overall growth rate
of the Rayleigh-Taylor instability, known as $\alpha$.
As a contrast to the above ILES picture, we note the
good validation results which have been achieved using FT/LES/SGS
(see Tables~\ref{table:RT-alpha} and \ref{table:RT-FT}) also
using DNS~\cite{MueAndSch1_09}.
The experiments RT give unique results, i.e., they are repeatable
(we allocate the 5-30\% experimental variation in RT $\alpha$ predominantly to
variation in recorded experimental conditions, in keeping with our
simulations which duplicate the experiment to experiment
variation in $\alpha$),
so that any solution (such as the FT/LES/SGS solution)
which agrees with these experiments is itself
numerically unique.

\begin{table}
\begin{center}
\caption{\label{table:RT-alpha}
Variabilities in $\alpha$
from a variety of experimental and numerical sources}
\begin{tabular}{|l|c|}
\hline
	\multicolumn{2}{|c|}{Experimental variabilities} \\
\hline
Experimental variability & 20\% \\
Due to experimental initial conditions & 5-30\% \\
\hline
\multicolumn{2}{|c|}{Numerical issues} \\
\hline
ILES to experiment discrepancy~\cite{DimYou04} & 100\% \\
ILES to ILES simulation discrepancy~\cite{DimYou04} & 50\% \\
Numerical variation
from transport coefficients~\cite{LimIweGli10,LimIweGli10a}& 5\%\\
FT/LES/SGS to experiment discrepancy~\cite{LimIweGli10a}& 5\% \\
\hline
\end{tabular}
\end{center}
\end{table}

\subsubsection{The RNG Framework and Nonuniqueness}
\label{sec:RNG-nonuniqueness}

The basic ideas relating the RNG framework to nonuniqueness
are explained in Sec.~\ref{sec:introduction}. Relevant to nonuniqueness is
the key RNG step of setting the parameters
for the essential variables.  (See Sec.~\ref{sec:RNG-steps}.)
We identify the essential variables,
tentatively and as a scientific judgment, as the
turbulent transport coefficients. These are set in RNG methodolgy by
comparison to experiment, and here, in lieu of an experiment,
as a coefficient of a model (Sec.~\ref{sec:models}). For the dynamic choice of
SGS terms, the coefficient is determined from the simulation itself in
Sec.~\ref{sec:dynamic}. The simulation is thus parameter free, and nonuniqueness
has been removed. The nonuniqueness originated in the unspecified 
turbulent transport and was removed by the RNG setting of the coefficients
of the essential variables, with the setting determined
by the dynamic method and thus uniquely.

\subsection{Verification and Validation}
\label{sec:vandv}

In~\cite{MelRaoKau12a}, we outline a validation/verification program for
LES in the high Re regime, based on the code FT/LES/SGS. 
The LES/SGS framework has no adjustable parameters. Our validation
is in the RT experimental regime.
We have validated the FT/LES/SGS code by
comparison to experimental measurements of the RT
growth rate $\alpha$, conducted at $\mathrm{Re} \sim 3.5 \times 10^4$.
This validation tests the transport coefficients, which in the
LES/SGS framework, are not adjustable, and which are much more 
sensitive in the RT experiments than in the RM experiments. Additionally,
through code comparison (referred to above), we compare FT to RAGE on RM
problems and rely on the RM validation of RAGE. Beyond the experimental
Re range of $3.5\times 10^4$, we employ an extrapolation, ie a
mathematical verification step, to Re values of $6\times 10^5$ to 
$6\times 10^7$ or higher. We observe that the transport coefficients
and the atomic level mixing CDFs display only a mild norm dependence
(a 10\% to 15\% effect) resulting from
a change  in the values of Re and are also a norm convergence under
mesh refinement, in a purely hydro study.

\subsubsection{V\&V for RT and RM Instabilities}
\label{sec:RT-RM}
We have conducted extensive studies of RT instabilities, as V\&V for the 
FT/LES/SGS algorithm. Expanding on the bottom two lines of 
Table~\ref{table:RT-alpha}, we summarize the principal results in 
Table~\ref{table:RT-FT}. We observe nearly perfect agreement between
simulation and experiment, within error bars, and accuracy sufficient
to distinguish between the distinct $\alpha$s of distinct experiments.
In this way, we show that the variation in $\alpha$ across multiple 
experiments is caused partly by initial conditions (the water channel
splitter plate experiments introduce significant noise) and partly
by changes in the fluid transport properties of the fluids 
\cite{LimIweGli10, LimIweGli10a}. The common belief that significant
long wave length noise present in the initial data explains the
factor of two discrepancy between experiment and the numerous ILES simulations,
has been shown to be false \cite{GliShaKam11}. The effect on $\alpha$ from the
long wave length perturbations in the initial condition 
of \cite{SmeYou87} was shown \cite{GliShaKam11} to be
$\pm5\%$.

\begin{table}
\caption{Comparison of FT simulation to experiment.
Discrepancy refers to
the comparison of results outside of uncertainty intervals, if any,
as reported.
\label{table:RT-FT}
}
\begin{center}
\begin{tabular}{|ccc|cc|c|}
\hline

Ref. & Exp. & Sim. Ref. & $\alpha_{\rm exp}$ & $\alpha_{\rm sim}$ & Discrepancy

\\
\hline
\cite{SmeYou87} &\#112 & \cite{LimIweGli10} & 0.052 & 0.055 & 6\%\\
\cite{SmeYou87} &\#105 & \cite{GliShaKam11} & 0.072 & $0.076\pm 0.004$ & 0\% \\
\cite{SmeYou87,Rea84} &10 exp. & \cite{GeoGliLi05}& 0.055-0.077 & 0.066 & 0\% \\

\cite{RamAnd04} & air-He & \cite{LiuGeoBo05} & 0.065-0.07 & 0.069 &0\% \\
\cite{Mue08} & Hot-cold & \cite{LimIweGli10,GliShaKam11} & $0.070\pm 0.011$ & 0.075 &  0\%\\
\cite{Mue08} & Salt-fresh & \cite{GliShaKam11} & $0.085 \pm 0.005$ & 0.084& 0\%\\
\hline
\end{tabular}
\end{center}
\end{table}

V\&V for micro observables is still an open research question, but partial
results have been obtained.
Convergence properties of the second moments of concentration and the
CDF (cumulative distribution function) for a RT instability are analyzed
in \cite{KamKauGli11,KauKamYu11,LimKamYu12}.
Convergence of the CDFs for the
joint concentration-temperature (micro) variables of an RM instability
were studied in \cite{MelRaoKau12a,MelRaoKau13}

\subsubsection{Comparison of Distinct SGS Model Formulations}
\label{sec:distinct_SGS}

The primary comparison issue addressed in this paper is the comparison
of a dynamic SGS algorithm, mostly of FT/LES/SGS
to ILES, based on published studies of RT simulations by the ILES method,
as discussed in Sec.~\ref{sec:numerically_observed}. The issue here is whether
to use SGS terms or not, whether they should be explicitly identified as
part of the simulation package or located implicitly
within the algorithm itself
in a non-transparent manner. Among the problems with ILES, and perhaps
related to its performance problems evident from Table~\ref{table:RT-alpha} of
Sec.~\ref{sec:numerically_observed} is the fact that the dimensionless
turbulent coefficients are ratios. Algorithms which seek to optimize
(and perhaps even succeed in optimizing)
the numerator and the denominator of a ratio
separately can still fail dramatically
in optimizing the ratio. Conceptually, the basis for optimization of
the numerator and the denominator is to reduce their size. But there is
no conceptual basis for optimization of the ratio. To the extent that
separate optimization of the numerator and denominator succeeds, the
ratio is the indeterminant expression 0/0.
It would seem that the ILES method, as it is generally explained,
does not offer a plan for such an optimization.

The methods used in the FT/LES/SGS  algorithm are a variant of the standard
dynamic SGS models, in that we average over a single mesh block,
rather than use an a localizing averaging kernel (filter), conventionally
extending over five mesh blocks. Additionally, in the numerical implementation,
the cell face averages are replaced by cell centered quantities, an 
approximation which reduces the accuracy of these terms to  first order.
We do not expect a major 
difference between the cell block averages used here (called 
an implicit filter) and the conventional extended filter for the dynamic 
SGS method. Aside from the simpler conceptual analysis of the mesh 
block averages , a feature exploited in the present paper, we prefer the mesh 
block averages as this method appears to be favorable when coupling
to reactive flows (turbulent combustion), with finite rate chemistry.
An assessment of this proposal will appear at a later time.

A separate and important question, 
but out of the scope of this paper, is to a quantitative comparison
of the variety of turbulence algorithms which do include explicit SGS terms.
The reader is referred to a variety of survey and review documents
e..g., \cite{MonKat00,LamFriGeu06} for this active research topic.
Dynamic SGS models \cite{Leo74,MoiSquCab91} 
are arguably the standard method to which others are
compared. We also mention the approximate deconvolution method (ADM), with a
recent contribution \cite{SchSolKle06}.
Comparisons involve specific flows and specific features of these
flows. In such comparisons, sometimes alternate methods excel.
Classical problems of turbulent flow fall into this category,
such as wall bounded turbulence, turbulent shock boundary layer interactions,
turbulent combustion, and turbulent particulate flow. Complex flows
are an open research area, for which a universal solution appears
unlikely.

\section{Discussion}
\label{sec:discussion}

We have re-examined the relation between LES and the RNG.
LES cannot be an RNG fixed point as it fails to be scale invariant.
However, the RNG expansion still applies, and based on this, we
derive a closed form expansion for the unclosed SGS terms.
To leading order, this expansion coincides with the Leonard
stress in the derivation of the dynamic SGS models. The full expansion
of the unclosed terms 
suggests a
higher order determination of the model coefficients.
For design of aircraft and for flow in pipes, lift and drag, the
important observables, are macro in nature. For reactive flows, such 
as combustion, micro observables of atomic mixing properties are
fundamental. They are the direct input to continuum level chemical
reaction rate laws. Accurate modeling of micro observables allows
finite rate LES chemistry, and elimination of flame structure models from
combustion simulations. Practical requirements of engineering simulations
also lead us to emphasize
the importance of LES and of compressible simulations.

Apparent LES nonuniqueness (both macro and micro) is known, 
as reviewed in~\S\ref{sec:numerical}.
The macro level apparent LES nonuniqueness
of RT simulations speaks for itself, even if 
its mathematical status in not known. 
It seems safe to speculate that the apparent LES nonuniqueness
at the micro (as opposed to the macro) level,
is more likely to survive future levels
of mesh refinement and careful control over simulation input data, and in this
sense it may be scientifically more fundamental.

We discuss the possible role of numerical
and physical modeling issues in the selection of a high Re limit point.
We have identified the RNG relevant variables
as the dimensionless parameters of the Euler equation.
The identification of these parameters as relevant is conventional
within RNG methods, but it is neither demonstrated by the
results of this paper nor is it essential to them.
RNG theories may also include dimensional equation
parameters, as with the mass of the electron in quantum field theory.
Some of the dimensionless turbulent transport parameters
could turn out to be irrelevant (decreasing more rapidly under
RNG iterations) and still parametrize non-unique solutions, to be
achieved by a stronger fractal or numerical algorithmic forcing.

There can be little doubt that a scientific
understanding of turbulent mixing and its dependence
on $\mathrm{Re}$ in the simple ``pure hydro'' example considered
here provides an indispensable foundation for the study
of mixing in the multi-physics context likely to prevail
in complex problems of engineering interest.

\appendix

\section{Elementary Formulas}
\label{sec:app}

Here we gather some elementary formulas, derived in detail for the
convenience of the reader.

\subsection{Covariance}
\label{sec:covariance-formulas}
We begin with the two-dimensional ($D = 2$) case of the grid $\mathcal{M}_1$,
which subdivides $[0,1] \times [0,1]$ into four squares,
and a particular face $f$ of the grid $\mathcal{M}_0$,
say $[0,1] \times \{0\}$,
which is subdivided by $\mathcal{M}_1$ into two segments.
A quantity $a = E_1 a$ that is $\mathcal{M}_1$-resolved
takes on a constant value on each of the two segments of $f$;
we represent $a$ as the two-dimensional vector $a = (a_1, a_2)^T$.
Such vectors form a two-dimensional real Hilbert space,
provided we adopt an inner product.
For the standard inner product $\langle \cdot, \cdot \rangle$,
the projection $E_0$ onto constant vectors
and the complementary projection $F_0 = I - E_0$
are given by
\begin{equation}
\label{eq:const}
\qquad
E_0 = \frac{1}{2}
\begin{pmatrix}
1 & \phantom{-}1 \\ 1 & \phantom{-}1
\end{pmatrix} \ , \qquad
F_0 = \frac{1}{2}
\begin{pmatrix}
\phantom{-}1 & -1 \\ -1 & \phantom{-}1
\end{pmatrix} \ .
\end{equation}

Now let $\rho = \left( \begin{smallmatrix}
\rho_1 & 0 \\ 0 & \rho_2 \end{smallmatrix} \right)$,
where $\rho_1$, $\rho_2 > 0$,
and consider the two-dimensional real Hilbert space
with weighted inner product $\langle \cdot, \cdot \rangle_\rho$
defined by
\begin{equation}
\langle a, b \rangle_\rho = \rho_1 a_1 b_1 + \rho_2 a_2 b_2
= \langle a, \rho b \rangle
\end{equation}
for vectors $a$ and $b$.
Next we compute the adjoint $A^{*\rho}$
of a matrix $A = \left( \begin{smallmatrix} a_{11}&a_{12}\\a_{21}&a_{22}
\end{smallmatrix} \right)$
with respect to $\langle \cdot, \cdot \rangle_\rho$.

For vectors $a = \widetilde{E}_1 a$ and $b = \widetilde{E}_1 b$,
\begin{equation}
\label{eq:adjoint}
\langle b, A a\rangle_\rho = \left\langle b, \rho A a \right\rangle
= \left\langle b, \rho A \rho^{-1} \rho a \right\rangle
= \left\langle \rho^{-1} A^T \rho b, \rho a \right\rangle
= \left\langle \rho^{-1} A^T \rho b, a \right\rangle_\rho \ ,
\end{equation}
where $A^T$ is the adjoint
with respect to the usual inner product,
i.e., the transpose of $A$.
Thus
\begin{equation}
A^{*\rho}
= \rho^{-1}\,A^T\,\rho
= \begin{pmatrix} a_{11} & (\rho_2/\rho_1)\,a_{21} \\
(\rho_1/\rho_2)\,a_{12} & a_{22} \end{pmatrix} \ .
\end{equation}
Therefore $A$ is self-adjoint in the $\rho$ inner product
if and only if $a_{12} = (\rho_2/\rho_1)\,a_{21}$.

Having determined the adjoint operation,
we verify that
\begin{align}
\widetilde{E}_0
&= \frac{1}{\rho_1 + \rho_2}
\begin{pmatrix} \rho_1 & \rho_2 \\ \rho_1 & \rho_2 \end{pmatrix}
= \frac{1}{\rho_1 + \rho_2} \begin{pmatrix} 1 \\ 1 \end{pmatrix}
\begin{pmatrix} \rho_1 & \rho_2 \end{pmatrix} \ , \\
\widetilde{F}_0
&= \frac{1}{\rho_1 + \rho_2}
\begin{pmatrix} \rho_2 & -\rho_2 \\ -\rho_1 & \rho_1 \end{pmatrix}
= \frac{1}{\rho_1 + \rho_2} \begin{pmatrix} -\rho_2 \\ \rho_1 \end{pmatrix}
\begin{pmatrix} -1 & 1 \end{pmatrix}
\end{align}
are the self-adjoint projection operators onto the subspace of constant vectors
and its complement with respect to $\langle \cdot, \cdot \rangle_\rho$,
the latter comprising states $\rho$-orthogonal to constant vectors.

With these preparations,
we compute that
\begin{equation}
\label{eq:E-evaluated}
\langle b, \widetilde{E}_0 a \rangle_\rho
= \left\langle \begin{pmatrix} b_1 \\ b_2 \end{pmatrix},
\frac{\rho_1\,a_1 + \rho_2\,a_2}{\rho_1 + \rho_2}
\begin{pmatrix} 1 \\ 1 \end{pmatrix} \right\rangle_\rho
= \frac{(\rho_1\,b_1 + \rho_2\,b_2)\,(\rho_1\,a_1 + \rho_2\,a_2)}
{\rho_1 + \rho_2}
\end{equation}
and
\begin{equation}
\label{eq:F-evaluated1}
\langle b, \widetilde{F}_0 a \rangle_\rho
= \left\langle \begin{pmatrix} b_1 \\ b_2 \end{pmatrix},
\frac{a_2 - a_1}{\rho_1 + \rho_2}
\begin{pmatrix} -\rho_2 \\ \rho_1 \end{pmatrix} \right\rangle_\rho
= \frac{\rho_1\,\rho_2}{\rho_1 + \rho_2}\,(b_2 - b_1)\,(a_2 - a_1) \ .
\end{equation}
Notice that
\begin{equation}
\langle b, \widetilde{F}_0 a \rangle_\rho
= \langle \widetilde{E}_1 b, \widetilde{F}_0 \widetilde{E}_1 a \rangle_\rho
= \langle b, \widetilde{E}_1 \widetilde{F}_0 a \rangle_\rho \ .
\end{equation}
Therefore the expression
$\widetilde{E}_0\left[b\,\widetilde{E}_1\widetilde{F}_0 a\right]$,
which is the leading order term
in the expansion~(\ref{eq:expansion})
of $\widetilde{a''\,b''}$ for $n = 0$,
takes the value~(\ref{eq:F-evaluated1}) on the face $f$ of $\mathcal{M}_0$.
This value is the product of finite differences.

We extend this analysis to three dimensions $(D = 3)$
for a face $f$ of an elementary $2^3$ grid.
The face is divided into a $2 \times 2$ grid with, say,
tangential coordinate directions $1$ and $2$.
The four values of a quantity $a$ on $f$
constitute the vector $(a_{11}, a_{12}, a_{21}, a_{22})^T$.
In terms of this notation,
the $\rho$-orthogonal projection onto constant vectors is
\begin{equation}
\widetilde{E}_0
= \frac{1}{\rho_{11} + \rho_{12} + \rho_{21} + \rho_{22}}
\begin{pmatrix} 1 \\ 1 \\ 1 \\ 1 \end{pmatrix}
\begin{pmatrix} \rho_{11} & \rho_{12} & \rho_{21} & \rho_{22} \end{pmatrix} \ .
\end{equation}
Explicit calculation then shows that
\begin{equation}
\label{eq:quad-explicit}
\langle b, \widetilde{F}_0 a \rangle_\rho
= \left(\sum_{i,j} \rho_{ij}\right)^{-1}
\sum_{(i,j) < (k,\ell)}
\rho_{ij} \rho_{k\ell}(b_{kl} - b_{ij})(a_{kl} - a_{ij}) \ ,
\end{equation}
where index pairs are ordered lexicographically,
so that there are six terms in the sum.
Each term is again the product of finite differences.

\subsection{Third-order cumulants}
\label{sec:cumulants-formulas}
To study the third-order cumulant, we examine the inner products
\begin{equation}
\label{eq:3-cum-inner-product}
\langle a, (\widetilde{F_0}b)(\widetilde{F_0}c)\rangle_\rho
\quad \text{and} \quad
\langle \widetilde{F_0}a, (\widetilde{F_0}b)(\widetilde{F_0}c)\rangle_\rho \ ,
\end{equation}
first for a single face in the $D=2$, $\mathcal{M}_1$ context.
By (\ref{eq:F-evaluated1}), we have
\begin{equation}
\label{eq:F-evaluated5}
\widetilde{F_0}a =
\frac{a_2 - a_1}{\rho_1 + \rho_2}
\begin{pmatrix} -\rho_2 \\ \rho_1 \end{pmatrix} \ .
\end{equation}
Thus
\begin{equation}
\label{eq:FF-evaluate1}
(\widetilde{F_0}b)( \widetilde{F_0}c) =
\frac{(b_2 - b_1)(c_2 - c_1)}{(\rho_1 + \rho_2)^2}
\begin{pmatrix} \rho_2^2 \\ \rho_1^2 \end{pmatrix}
\end{equation}
and
\begin{equation}
\label{eq:FF-evaluate2}
\langle a, (\widetilde{F_0}b)(\widetilde{F_0}c)\rangle_\rho =
(a_1\rho_1 + a_2\rho_2)\frac{\rho_1\rho_2}{\rho_1 +\rho_2}
(b_2 - b_1)(c_2 - c_1) \ ,
\end{equation}
\begin{equation}
\label{eq:FF-evaluate3}
\langle \widetilde{F_0}a, (\widetilde{F_0}b)(\widetilde{F_0}c)\rangle_\rho =
(a_2 - a_1)(b_2 - b_1)(c_2 - c_1)
\frac{\rho_1\rho_2(\rho_2-\rho_1)}{(\rho_1+\rho_2)^2} \ .
\end{equation}

Again we extend this analysis to three dimension, for a face of an
elementary $2^3$ grid. Consider a face parallel to the $x,y$ plane.
We introduce the $x$ direction average operator $E_{0,x}$, which projects 
onto states constant in the $x$ direction, and its orthogonal compliment,
$F_{0,x}$, which projects onto states orthogonal to $x$ direction constants.
Similarly we introduce $E_{0,y}$ and $F_{0,y}$. We have
\begin{equation}
\label{eq:xy_expansion}
I = E_{0,x}+F_{0,x}=E_{0,x}E_{0,y}+E_{0,x}F_{0,y}
+F_{0,x}E_{0,y}+F_{0,x}F_{0,y} \ .
\end{equation}
We substitute this identity
into the cumulant expression, in front of each of $a,b,c$.
Only terms with at least on $F$ for each factor are nonzero, and so there are
27 of these. We do not present detailed formulas for the 27 terms, but we note
that each $F$ introduces a first difference operator. Thus each of the 27
terms is a monomial in first differences applied at least once to each of
$a,b,c$. It has cubic, up to 6 power of first differences within the 
27 terms.

\bibliography{refs}
\bibliographystyle{siam}

\end{document}